\documentclass{article}
\usepackage{spconf,amsmath,graphicx,hyperref}

\usepackage{amsmath,amssymb}
\usepackage{comment}
\usepackage{float}
\usepackage{subcaption}
\usepackage{algorithm}
\usepackage{algorithmic}

\usepackage{enumitem}
\setlist[itemize]{itemsep=0pt, parsep=0pt, topsep=0pt}
\title{LATENT-SPACE METRICS FOR COMPLEX-VALUED VAE OUT-OF-DISTRIBUTION DETECTION UNDER RADAR CLUTTER}
\name{
     Y. A. Rouzoumka$^{1,2}$\thanks{Part of this work was supported by ANR-ASTRID NEPTUNE 3 (ANR-23-ASM2-0009).} \qquad
    E. Terreaux$^{1}$ \qquad
    C. Morisseau$^{1}$ \qquad
   J.-P. Ovarlez$^{1,2}$ \quad
   C. Ren$^{2}$
}
\address{
    $^{1}$ DEMR, ONERA, Universit\'e Paris-Saclay, 91120 Palaiseau, France \\
    $^{2}$ SONDRA, CentraleSup\'elec, Universit\'e Paris-Saclay, 91190 Gif-sur-Yvette, France.   
}

\begin{document}
%
\maketitle
\begin{abstract}
We investigate complex-valued Variational AutoEncoders (CVAE) for radar Out-Of-Distribution  (OOD) detection in complex radar environments. We proposed several detection metrics: the reconstruction error of CVAE (CVAE-MSE), the latent-based scores (Mahalanobis, Kullback-Leibler divergence (KLD)), and compared their performance against the classical ANMF-Tyler detector (ANMF-FP). The performance of all these detectors is analyzed on synthetic and experimental radar data, showing the advantages and the weaknesses of each detector. 
\end{abstract}


\section{Introduction}
\label{sec:intro}


Classical adaptive radar detectors (Kelly’s detector, AMF, ANMF) rely on Gaussian and SIRV clutter models and may degrade under non-Gaussian plus additive thermal noise backgrounds \cite{4104190,Robey1992ACA,301849}. We therefore pursue a data-driven alternative based on complex-valued VAEs and latent-space OOD scores.

In recent years, data-driven approaches have emerged to alleviate the need for precise clutter modeling. Among them, VAEs \cite{Kingma_2019} have demonstrated promising capabilities for anomaly and OOD detection in diverse applications, including radar detection \cite{RouzoumkaICASSP}, speech enhancement \cite{complexreccurrxie}, medical imaging \cite{marimont2020}, industrial monitoring \cite{MSWPNM2021}, and acoustic signal analysis \cite{BS2023}. These models learn a latent representation of the training data and use reconstruction or probabilistic criteria to detect deviations. Despite their effectiveness, most VAE-based detectors operate in the real domain and often treat complex-valued radar data by separating real and imaginary components into distinct channels.

Recent advances in Complex-Valued Neural Networks (CVNNs) have shown the benefits of directly modeling complex-valued signals \cite{trabelsi2018deepcomplex, barrachina:hal-03771786}. In radar signal processing, complex-valued models are particularly well-suited due to the inherent phase information contained in the data. Complex-Valued VAEs (CVAEs) extend this paradigm, enabling phase-aware latent modeling and better structural encoding of radar signals. 

However, while CVAEs offer promising avenues for radar detection, the literature lacks a comprehensive evaluation of how different OOD detection strategies behave in the complex latent space. Most existing works rely on simple reconstruction error metrics (e.g., MSE) in the observation domain, without exploiting the statistical richness of the learned latent distribution. Moreover, it remains unclear whether latent-based scores such as Mahalanobis distance \cite{9053206}, KLD, or magnitude-based norms provide robust alternatives for radar detection, especially under challenging noise conditions (e.g., compound clutter or low SNR regimes).

We present the study of complex-valued latent-space OOD scores learned by a CVAE for radar, and position them against a classical detector. Our contributions are (i) benchmark of CVAE latent-space OOD scores (Mahalanobis and empirical KLD) versus the reconstruction error (CVAE-MSE) \emph{and} a strong classical baseline (ANMF-Tyler built with Tyler estimate) \cite{rouzoumka2025complexvae}, all calibrated at $P_{fa}=10^{-2}$, (ii) evaluate across cGN+AWGN, cCGN+AWGN, and experimental CSIR data, reporting both the mean over cells (excluding Doppler~0) and the specific Doppler-0 cell, (iii) analyze robustness vs.\ SNR: (1) CVAE-MSE leads on synthetic data; (2) ANMF-FP is particularly effective under compound-Gaussian clutter and on real data; (3) Hermitian Mahalanobis mitigates Doppler-0 degradation relative to CVAE-MSE, (iv) KLD is brittle in practice and (iv) give guidelines on selecting reconstruction, latent, or classical scores depending on clutter realism and cell-specific behavior (e.g., Doppler~0).

\indent \textit{Notations}: Matrices are in bold and capital, vectors in bold. For any matrix $\mathbf{A}$ or vector, $\mathbf{A}^T$ is the transpose of $\mathbf{A}$, $\mathbf{A}^*$ the (elementwise) complex conjugate, and $\mathbf{A}^H=(\mathbf{A}^*)^T$ the Hermitian transpose of $\mathbf{A}$. $\mathbf{I} $ is the identity matrix. $\mathcal{N}(\boldsymbol{\mu},\boldsymbol{\Gamma})$ and $\mathcal{CN}(\boldsymbol{\mu},\boldsymbol{\Gamma})$ are respectively real and complex circular Normal distribution of mean $\boldsymbol{\mu}$ and covariance matrix $\boldsymbol{\Gamma}$. The matrix operator $ \boldsymbol{\mathcal{T}}(.)$ is the Toeplitz matrix operator $\rho \rightarrow \left\{\boldsymbol{\mathcal{T}}(\rho)\right\}_{i,j} = \rho^{|i-j|}$. The symbols $\odot$ and $\oslash$ denote the Hadamard element-wise product and division, respectively. On vectors, the operator $^\circ$ acts on each of its components (power, log, modulus).

\section{Related Work}
\label{sec:related}

VAEs are widely used for unsupervised OOD detection. A common baseline flags samples with large reconstruction error (or reconstruction probability) as anomalies~\cite{an2015variational}. Beyond reconstruction, latent-space scores are often employed: the Kullback-Leibler divergence between the posterior $q(z|x)$ and the prior $p(z)$~\cite{sohn2015cvae}, and the Mahalanobis distance computed from deep features or latent embeddings~\cite{lee2018simple}. These methods remain foundational in recent comparisons of VAE-based anomaly/OOD detectors~\cite{nguyen2024vae}. 


Complex-valued deep learning natively matches radar data. Foundational work on complex-valued networks established the building blocks and benefits for structured signals~\cite{trabelsi2018deepcomplex}. Despite this momentum, complex-valued VAEs applied to radar remain scarce. To the best of our knowledge, prior work in adjacent signal-processing domains (e.g., speech enhancement) demonstrates the practicality of complex-valued VAEs, but radar-specific CVAE detectors are still under-explored; our study fills this gap by benchmarking latent-space criteria in complex-valued settings.

Latent-space scoring has a long pedigree: the Mahalanobis distance~\cite{mahalanobis1936} adapted to deep features offers a class-conditional statistical deviation measure~\cite{lee2018simple}, while the VAE’s KLD term provides a natural posterior-prior mismatch signal~\cite{sohn2015cvae}. However, most of these investigations focus on \emph{real-valued} latents; their behavior in complex-valued generative models (and in radar specifically) remains comparatively underexplored, motivating the contributions of this work.





\section{Methodology}
\label{sec:methodology}

This section presents our comprehensive framework for evaluating latent space detection methods in complex-valued VAEs. We first describe the complex-valued VAE architecture and its adaptations for radar signal processing \cite{rouzoumka2025complexvae}, then provide detailed mathematical formulations and methodological developments for our two primary contributions: KL divergence-based detection and Hermitian Mahalanobis distance in complex latent spaces.

\subsection{Complex-Valued VAE Architecture}
\label{ssec:cvae_arch}

Our approach builds upon the complex-valued VAE architecture specifically designed for radar signal processing applications. Unlike conventional real-valued VAEs that require separate processing of real and imaginary components, our model operates directly in the complex domain, preserving crucial phase information inherent in radar signals.

The encoder maps high-dimensional complex radar signals $\mathbf{x} \in \mathbb{C}^N$ into a lower-dimensional latent representation through complex-valued convolutional layers with complex batch normalization and $\mathbb{C}$ReLU activation functions. The encoder outputs three key parameters: the mean vector $\boldsymbol{\mu} \in \mathbb{C}^q$, the variance vector $\boldsymbol{\sigma}^{\circ 2} \in \mathbb{R}^{q+}$,
 and the pseudo-variance term $\boldsymbol{\delta} \in \mathbb{C}^q$, which allows greater flexibility in modeling the latent space.

\begin{figure}[htb]
\centering
\includegraphics[width=1.\columnwidth]{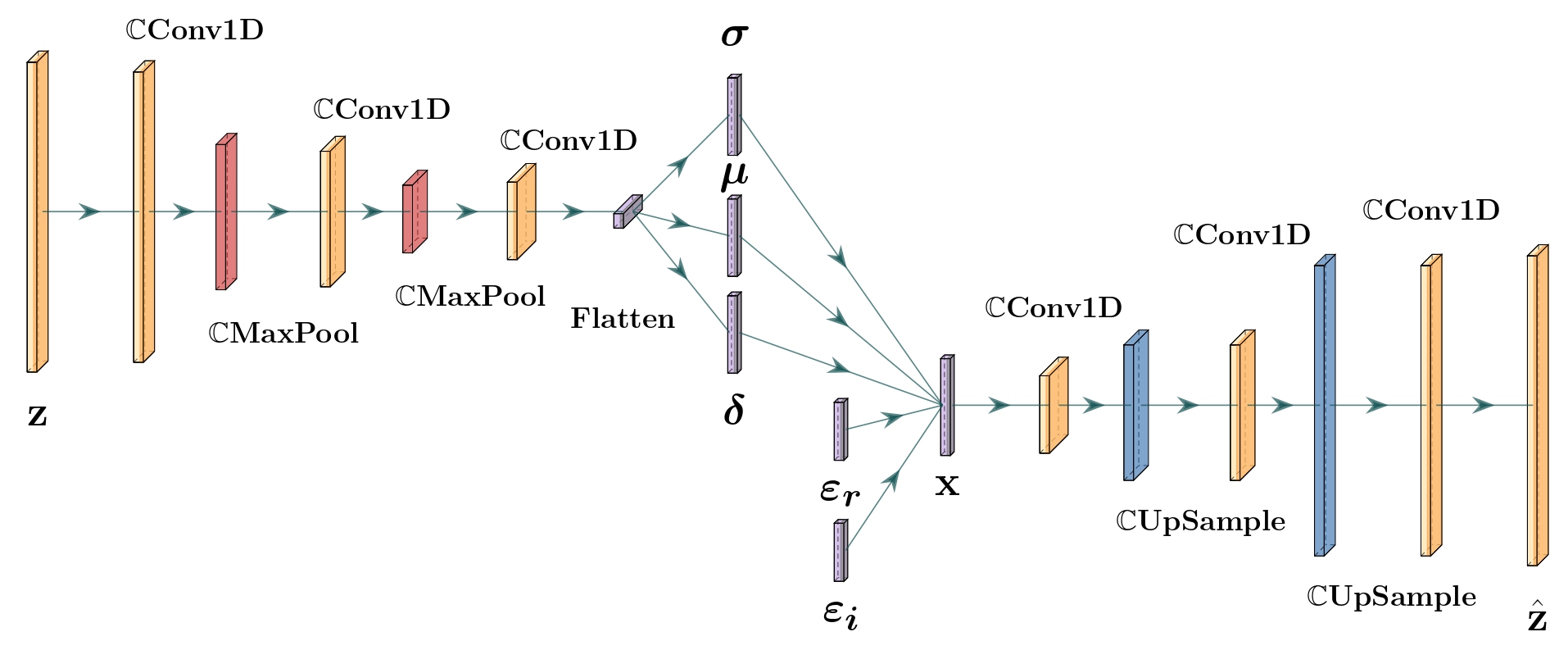}
\caption{Complex-Valued VAE network architecture}
\label{fig:vae}
\end{figure}

To enable efficient sampling in the complex latent space while maintaining differentiability, we introduce a novel reparameterization trick. Given the estimated parameters $\boldsymbol{\mu}$, $\boldsymbol{\sigma}$, and $\boldsymbol{\delta}$, the latent variable $\mathbf{z}$ is sampled as follows \cite{nakashika20}:
\begin{equation}
    \mathbf{z} = \boldsymbol{\mu} + \boldsymbol{k}_r \odot \boldsymbol{\epsilon}_r + i\, \boldsymbol{k}_i \odot \boldsymbol{\epsilon}_i,
\end{equation}
where $\left\{\begin{array}{l} 
    \boldsymbol{k}_r = \frac{1}{\sqrt{2}}\left(\boldsymbol{\sigma} + \boldsymbol{\delta}\right) \oslash \left(\boldsymbol{\sigma} + \mathrm{Re}(\boldsymbol{\delta})\right)^{\circ \frac{1}{2}}, \\
    \boldsymbol{k}_i = \frac{1}{2} \left(\boldsymbol{\sigma}^{\circ 2} - \left(|\boldsymbol{\delta}|^\circ\right)^{\circ 2} \right)^{\circ \frac{1}{2}} \oslash \left(\boldsymbol{\sigma} + \mathrm{Re}(\boldsymbol{\delta})\right)^{\circ \frac{1}{2}}\, .
\end{array}\right.$

Here, $\boldsymbol{\epsilon}_r$ and $\boldsymbol{\epsilon}_i$ are identically and independently distributed according to standard Gaussian noise vectors $\mathcal{N}(\mathbf{0}, \mathbf{I})$. 


The decoder reconstructs the radar signal $\mathbf{x} \in \mathbb{C}^N$ from the latent variable $\mathbf{z}$ using transposed complex convolutions, complex batch normalization, and $\mathbb{C}$ReLU activations. It mirrors the encoder in reverse to upsample and restore the original signal structure.

The model is trained by minimizing the complex-valued ELBO:
\begin{equation}
    \mathcal{L}_{\text{CVAE}} = \mathcal{L}_{\text{rec}} + \beta \, \mathcal{D}_{\text{KL}}\, ,
\end{equation}
where the reconstruction loss is $\mathcal{L}_{\text{rec}} = \|\mathbf{x} - \hat{\mathbf{x}}\|^2$ and the KL divergence for complex latent spaces against a standard circular complex Gaussian prior is:
\begin{equation}
    \mathcal{D}_{\text{KL}} = \|\boldsymbol{\mu}\|^2 + \mathbf{1}_q^T \Big(\boldsymbol{\sigma} - \frac{1}{2} \log^\circ\left(\boldsymbol{\sigma}^{\circ 2} - \left(|\boldsymbol{\delta}|^\circ \right)^{\circ 2}\right) \Big)\, .
\end{equation}

\subsection{Detection Scores Considered}
We compare three scores computed with a complex-valued VAE trained on clutter: 
(i) a reconstruction-based baseline (CVAE\_MSE), and two latent-space scores:
(ii) an empirical complex KL divergence against the marginal $H_0$, and 
(iii) a Hermitian Mahalanobis distance in the complex latent space.

\subsubsection{Reconstruction-Based Detection (CVAE\_MSE)}
\label{sssec:vae_mse}
Our baseline follows the standard CVAE paradigm: targets are identified through reconstruction error defined as $S_{\text{MSE}}(\mathbf{x})=\|\mathbf{x}-\hat{\mathbf{x}}\|^2$ 
with $\mathbf{x},\hat{\mathbf{x}}\in\mathbb{C}^N$.
The binary decision uses a threshold $\lambda_{\text{MSE}}$ calibrated on clutter ($H_0$) to meet a prescribed PFA:
\begin{equation}
S_{\text{MSE}}(\mathbf{x})\underset{H_0}{\overset{H_1}{\gtrless}}\lambda_{\text{MSE}}\, .
\end{equation}

\subsubsection{KLD for Empirical $H_0$ complex Gaussian distribution}

Rather than enforcing a theoretical prior $\mathcal{CN}(\mathbf{0},\mathbf{I})$, we estimate in latent space an empirical complex Gaussian for clutter:
\(
\mathcal{CN}(\hat{\boldsymbol{\mu}}_0,\hat{\boldsymbol{\Sigma}}_0,\hat{\boldsymbol{\Delta}}_0)
\),
where $\hat{\boldsymbol{\Sigma}}_0$ captures second-order covariance and 
$\hat{\boldsymbol{\Delta}}_0$ (diagonal) models residual non-circularity. 
Deviations of $q_{\boldsymbol{\phi}}(\mathbf{z}\!\mid\!\mathbf{x})$ from this law are used for detection.

For clutter samples $\mathbf{x}_i$ the encoder outputs $(\boldsymbol{\mu}_i,\boldsymbol{\sigma}_i,\boldsymbol{\delta}_i)$.
By setting $k_{i,\ell}=\max(\sigma_{i,\ell}^2-|\delta_{i,\ell}|^2,\varepsilon)$ (stability clip), we can define the per-sample \emph{circular} covariance
$\boldsymbol{\Sigma}_\mathrm{enc}(\mathbf{x}_i)=\mathrm{diag}(k_{i,1},\ldots,k_{i,q})$ and :
\begin{equation}
\hat{\boldsymbol{\mu}}_0 = \displaystyle\frac{1}{N_c}\sum_{i=1}^{N_c}\boldsymbol{\mu}_i,
\,\,\, \hat{\boldsymbol{\Sigma}}_{0} = \frac{1}{N_c}\sum_{i=1}^{N_c}\boldsymbol{\Sigma}_\mathrm{enc}(\mathbf{x}_i)\, ,
\label{eq:totcov}
\end{equation}
\begin{algorithm}[htbp]
\caption{Empirical KL Detection}
\label{alg:empirical_kl_detection_totalcov}
\begin{algorithmic}[1]
\REQUIRE Encoder $\texttt{Enc}$, clutter set $\mathcal{D}_{0}$, test sample $\mathbf{x}$, target PFA $\alpha$
\STATE For each $\mathbf{x}_i\!\in\!\mathcal{D}_{0}$, get $(\boldsymbol{\mu}_i,\boldsymbol{\sigma}_i,\boldsymbol{\delta}_i)\!\leftarrow\!\texttt{Enc}(\mathbf{x}_i)$; set $\boldsymbol{\Sigma}_1(\mathbf{x}_i)=\mathrm{diag}(\boldsymbol{k}_i)$ with $k_{i,\ell}=\max(\sigma_{i,\ell}^2-|\delta_{i,\ell}|^2,\varepsilon)$ and $\boldsymbol{\Delta}_1(\mathbf{x}_i)=\mathrm{diag}(\boldsymbol{\delta}_i)$
\STATE Estimate $(\hat{\boldsymbol{\mu}}_0,\hat{\boldsymbol{\Sigma}}_0)$ via \eqref{eq:totcov} and $\hat{\boldsymbol{\Delta}}_0=\mathrm{diag}(\tfrac{1}{N_c}\sum_i \boldsymbol{\delta}_i)$
\STATE Calibrate $\lambda_{\mathrm{KL}}=\mathrm{Percentile}\!\big(\{S_{\mathrm{KL}}(\mathbf{x}_i)\},1-\alpha\big)$ on $\mathcal{D}_{0}$
\STATE On $\mathbf{x}$: compute $(\boldsymbol{\mu},\boldsymbol{\sigma},\boldsymbol{\delta})\!\leftarrow\!\texttt{Enc}(\mathbf{x})$, form $\boldsymbol{\Sigma}_{enc}(\mathbf{x})$ and $\boldsymbol{\Delta}_{enc}(\mathbf{x})$, then $S_{\mathrm{KL}}(\mathbf{x})$ via \eqref{eq:skl}
\STATE Decide $H_1$ iff $S_{\mathrm{KL}}(\mathbf{x})>\lambda_{\mathrm{KL}}$
\end{algorithmic}
\end{algorithm}

We also estimate a diagonal pseudo-covariance as the average encoder output:
\(
\hat{\boldsymbol{\Delta}}_0=\frac{1}{N_c}\sum_i \mathrm{diag}\big(\boldsymbol{\delta}_i\big).
\)

For any test posterior density
$\mathcal{CN}(\boldsymbol{\mu}_{\text{enc}},\boldsymbol{\Sigma}_{\text{enc}},\boldsymbol{\Delta}_{\text{enc}})$,
the KL distance to the empirical null
$\mathcal{CN}(\hat{\boldsymbol{\mu}}_0,\hat{\boldsymbol{\Sigma}}_0,\hat{\boldsymbol{\Delta}}_0)$ is

\begin{align}
\label{eq:kl_closed}
D_{\mathrm{KL}}
&= \frac{1}{2}\Big[
\log\frac{\det\boldsymbol{K}_0}{\det\boldsymbol{K}_{\text{enc}}}
+\mathrm{tr}\big(\boldsymbol{K}_0^{-1}\boldsymbol{K}_{\text{enc}}\big) \\
&+(\boldsymbol{m}_0-\boldsymbol{m}_{\text{enc}})^{H}\boldsymbol{K}_0^{-1}(\boldsymbol{m}_0-\boldsymbol{m}_{\text{enc}})
-2q\Big]. \nonumber
\end{align}

where the augmented mean and covariance are
\[
\boldsymbol{m}_i=\begin{bmatrix}\boldsymbol{\mu}_i\\ \boldsymbol{\mu}_i^{*}\end{bmatrix},\qquad
\boldsymbol{K}_i=\begin{bmatrix}\boldsymbol{\Sigma}_i & \boldsymbol{\Delta}_i\\
\boldsymbol{\Delta}_i^{*} & \boldsymbol{\Sigma}_i^{*}\end{bmatrix},\quad i\in\{\text{enc},0\}.
\]

The detection score is then $S_{\mathrm{KL}}(\mathbf{x})\underset{H_0}{\overset{H_1}{\gtrless}}\lambda_{\mathrm{KL}}$ where 
\begin{equation}
\label{eq:skl}
S_{\mathrm{KL}}(\mathbf{x})\equiv D_{\mathrm{KL}}\big(q_{\phi}(\mathbf{z}\!\mid\!\mathbf{x})\,\|\,\mathcal{CN}(\hat{\boldsymbol{\mu}}_0,\hat{\boldsymbol{\Sigma}}_0,\hat{\boldsymbol{\Delta}}_0)\big)\, ,
\end{equation}

\subsubsection{Mahalanobis Distance in Complex Latent Space}
\label{ssec:mahalanobis_contribution}

For $\mathbf{z}\in\mathbb{C}^q$ with mean $\boldsymbol{\mu}_{\mathrm{ref}}$ and Hermitian covariance $\boldsymbol{\Sigma}_{\mathrm{ref}}$, we define the Mahalanobis distance as:
\begin{equation}
\label{eq:maha_sq}
S_{\mathrm{Maha}}(\mathbf{z})=\left(\mathbf{z}-\boldsymbol{\mu}_{\mathrm{ref}}\right)^H \, \boldsymbol{\Sigma}_{\mathrm{ref}}^{-1} \, \left(\mathbf{z}-\boldsymbol{\mu}_{\mathrm{ref}}\right)\in\mathbb{R}_+,
\end{equation}
which is unitarily invariant and fully accounts for complex correlations, and where the  following statistical parameters are estimated from $\{\mathbf{z}_n\}_{n=1}^N$:
\begin{equation}
\hat{\boldsymbol{\mu}}_{\mathrm{ref}}=\tfrac{1}{N}\sum_{n}\mathbf{z}_n\,, \, \, 
\hat{\boldsymbol{\Sigma}}_{\mathrm{ref}}=\tfrac{1}{N-1}\sum_{n}(\mathbf{z}_n-\hat{\boldsymbol{\mu}}_{\mathrm{ref}})(\mathbf{z}_n-\hat{\boldsymbol{\mu}}_{\mathrm{ref}})^H,
\label{eq:Mahalanobisparameters}
\end{equation}
and the decision uses a percentile threshold on $S_{\mathrm{Maha}}$ calibrated on $H_0$ to meet PFA~$\alpha$.

For each score, $\lambda$ is set on an independent $H_0$ validation set to reach $\mathrm{PFA}=\alpha$:
$\ \lambda=\mathrm{Percentile}\big(S_{\text{clutter}},1-\alpha\big)$.

We report Probability of Detection (PD) versus SNR, averaged over multiple Monte-Carlo trials, across Gaussian, compound Gaussian, and real data.

\begin{algorithm}[H]
\caption{Mahalanobis Detection}
\label{alg:mahalanobis_detection_short}
\begin{algorithmic}[1]
\REQUIRE Encoder $\texttt{Enc}$, clutter set $\mathcal{D}_{0}$, test sample $\mathbf{x}$, target PFA $\alpha$
\STATE Build $\mathbf{z}_i\!\leftarrow\!\texttt{Enc}(\mathbf{x}_i)$ for all $\mathbf{x}_i\in\mathcal{D}_0$
\STATE Estimate $(\hat{\boldsymbol{\mu}}_{\mathrm{ref}},\hat{\boldsymbol{\Sigma}}_{\mathrm{ref}})$ via \eqref{eq:Mahalanobisparameters}
\STATE Calibrate $\lambda_{\mathrm{Maha}}=\mathrm{Percentile}\!\big(\{S_{\mathrm{Maha}}(\mathbf{z}_i)\},1-\alpha\big)$
\STATE On $\mathbf{x}$: get $\mathbf{z}_{\mathrm{test}}$, compute $S_{\mathrm{Maha}}(\mathbf{z}_{\mathrm{test}})$ via \eqref{eq:maha_sq}
\STATE Decide $H_1$ iff $S_{\mathrm{Maha}}(\mathbf{z}_{\mathrm{test}})>\lambda_{\mathrm{Maha}}$
\end{algorithmic}
\end{algorithm}



\begin{figure*}[htb]
\centering
\begin{subfigure}{0.66\columnwidth}
\includegraphics[width=\linewidth,clip]{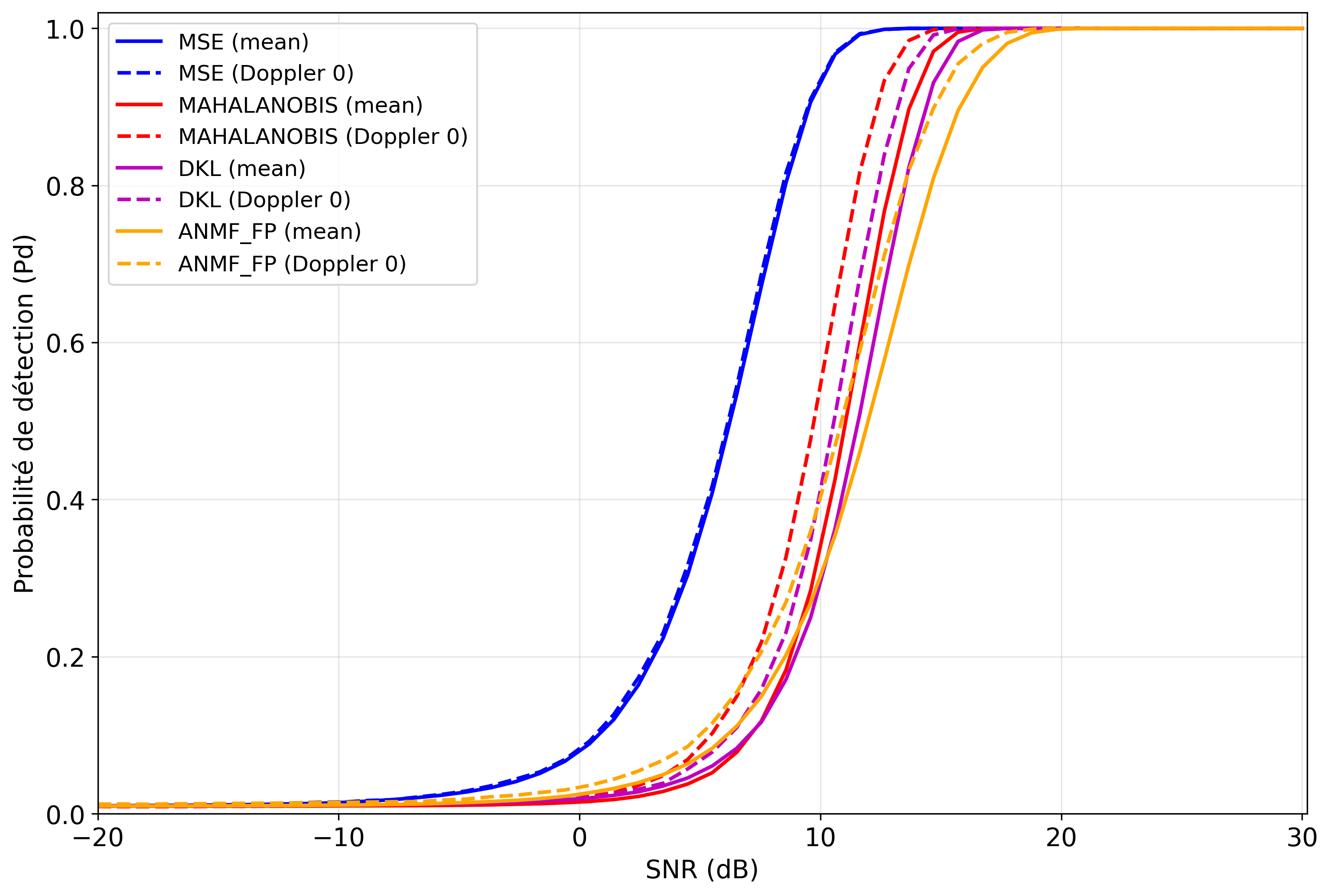}
\caption{cGN + AWGN}
\end{subfigure}
\hfill \begin{subfigure}{0.66\columnwidth}
\includegraphics[width=\linewidth,clip]{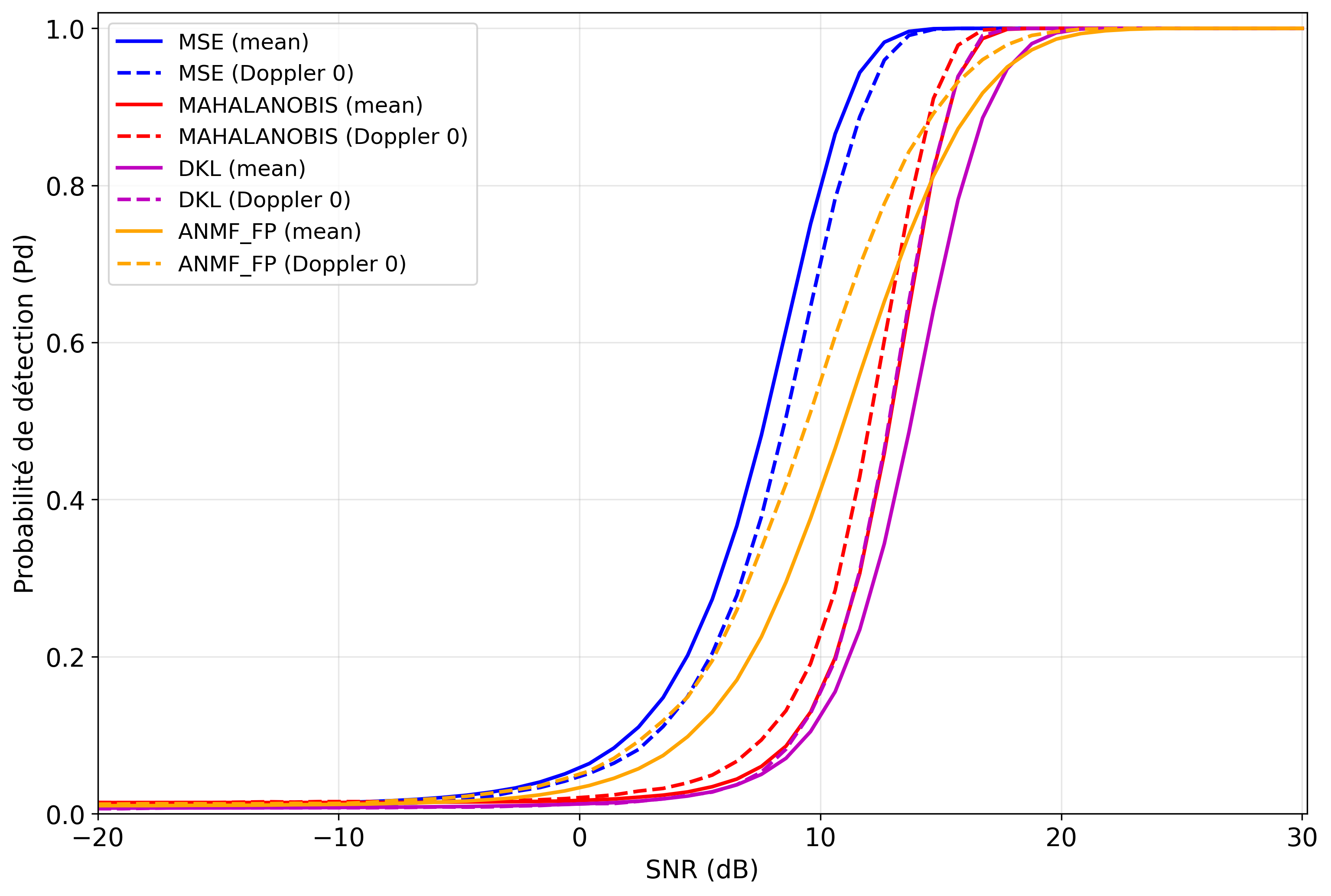}
\caption{cCGN + AWGN}
\end{subfigure}
\hfill \begin{subfigure}{0.66\columnwidth}
\includegraphics[width=\linewidth,clip]{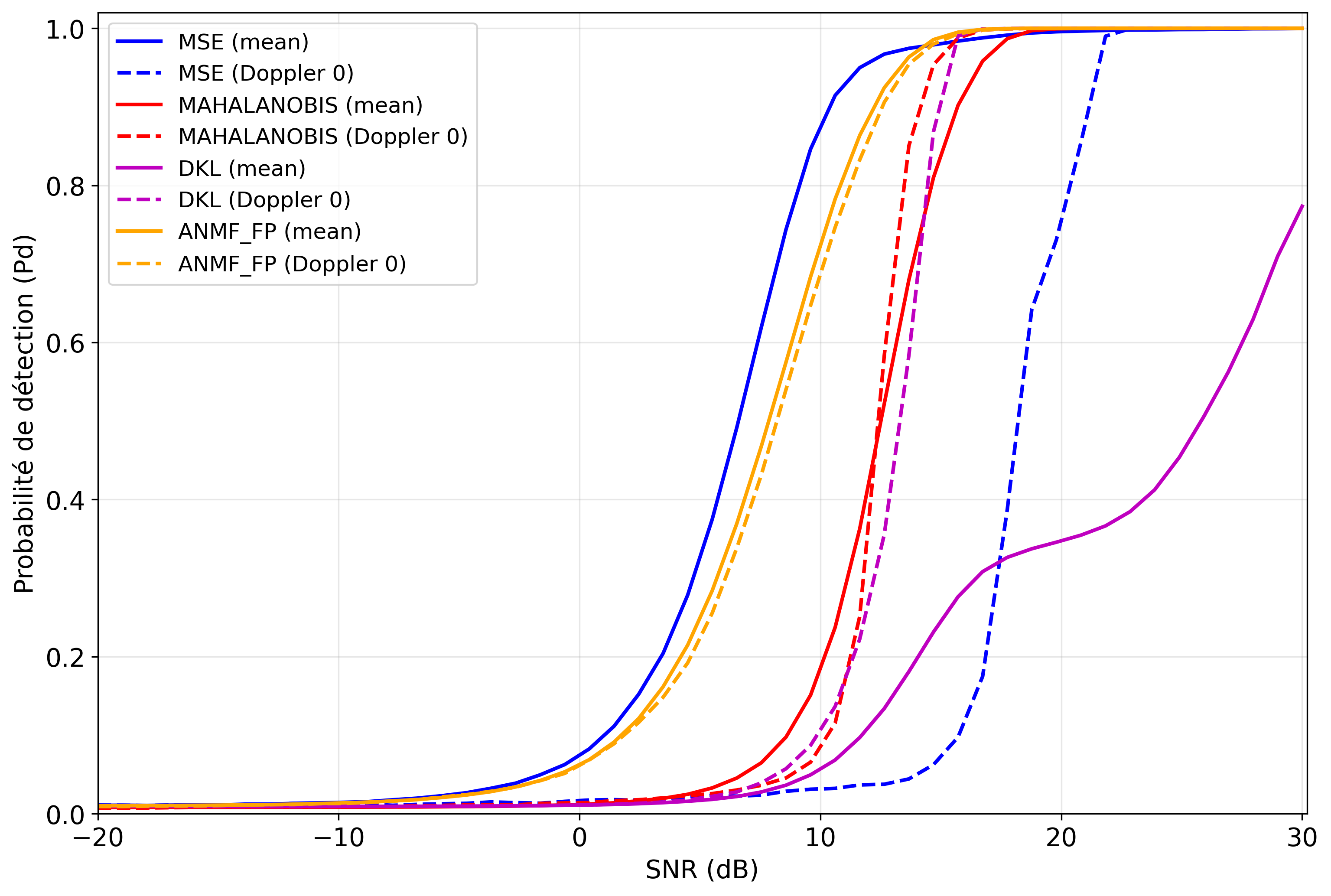}
\caption{CSIR data}
\end{subfigure}

\caption{Detection performance under different noise configurations, ($P_{fa}= 10^{-2}$).} 
 \label{fig:cpdsnr}
\end{figure*}

\section{Results and Discussion}
\label{sec:results}

We assess Probability of Detection $P_d$ vs.\ SNR at fixed $P_{fa}=10^{-2}$ under three conditions: \emph{(a)} cGN+AWGN, \emph{(b)} cCGN+AWGN, and \emph{(c)} real CSIR data.
Figures report solid lines for the mean over all cells except Doppler~0, and dashed lines for the Doppler-0 cell.

\subsection{Signal and Noise Characteristics}

The target echo is modeled as $\alpha = \sqrt{\text{SNR}} \, \exp\left(2i\pi \phi\right) / \sqrt{m}$, where $\phi \in [0,1]$, and the steering vector is given by $\mathbf{p} = \left(1, \exp\left(2i\pi d / m\right), \ldots, \exp\left(2i\pi d (m-1)/m\right) \right)^T$ for $m = 16$ bins, where \( d \) representing the target normalized Doppler bin index. The clutter covariance matrix follows $\boldsymbol{\Sigma}_c = \boldsymbol{\mathcal{T}}(\rho)$ with $\rho = 0.5$, while texture components $\tau$ and $\tau_k$ are sampled from a Gamma distribution $\Gamma(\mu, 1/\mu)$ with $\mu = 1$. For adaptive detectors, the covariance estimation uses SCM and Tyler estimators with $K = 2\,m$ independent secondary samples.

\begin{figure}[htb]
    \centering
    \begin{subfigure}{0.5\linewidth}
        \centering
        \includegraphics[width=\linewidth, clip]{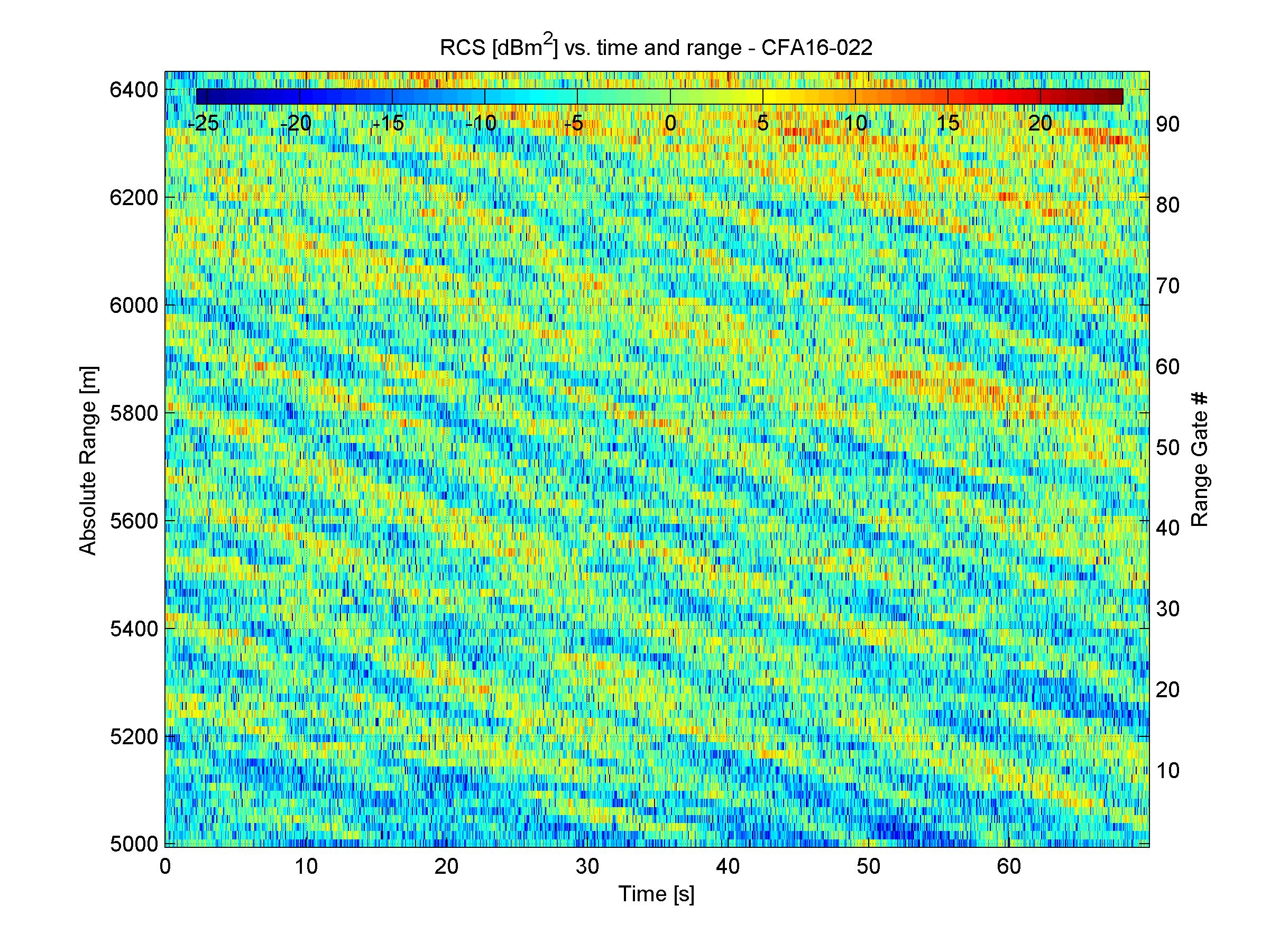}
    \end{subfigure}%
    \hfill
    \begin{subfigure}{0.5\linewidth}
        \centering
        \includegraphics[width=\linewidth,clip]{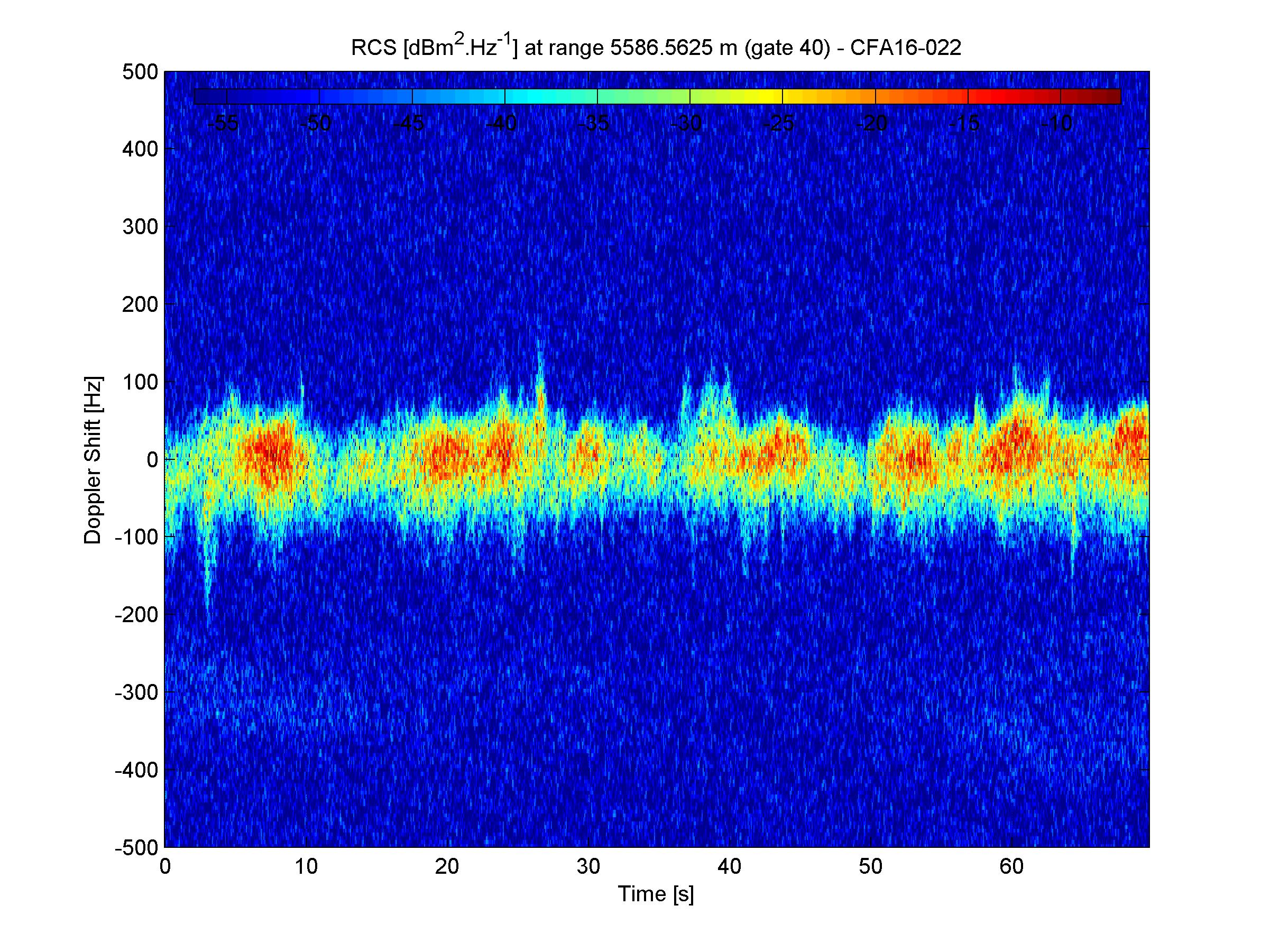}
    \end{subfigure}
    \caption{On the left: range-slow time map. On the right: Doppler-pulse spectrogram for range bin 5605.}
    \label{fig:carteCSIR}
\end{figure}

Concerning experimental data, we used the South Africa CSIR (Council for Scientific and Industrial Research) dataset, which comprises radar echoes from sea clutter and naval targets, collected during two campaigns: in 2006 at the Overberg Test Range with the Fynment radar, and in 2007 on Signal Hill with an experimental monopulse radar. It spans diverse waveforms, azimuths, ranges, and environmental conditions (see Fig.~\ref{fig:carteCSIR}).

\subsection{CVAE Training Configuration}
\label{ssec:trainprocess}
The VAE is trained on clutter-plus-noise Doppler profiles for each noise scenario. 
The dataset $\mathcal{D}_{H_0}$ consists of $15{,}000$ samples, two-thirds of which are allocated for training and the remaining third for validation. 
Training is performed over $50$ epochs using the Adam optimizer~\cite{Adam} with a learning rate of $10^{-3}$. The loss function $\mathcal{L}_{\text{CVAE}}$ incorporates a regularization parameter of $\alpha = 10^2$ for the CVAE-MSE and $\alpha = 10^{-3}$. The latent space dimension is set to $12$ for the VAE-MSE and to $32$ for the Mahalanobis and KLD latent detectors. Once trained, detection is performed using the reconstruction loss $\mathcal{L}_{\text{rec}}$, with $P_{fa} = 10^{-2}$, determined from a validation set of $5{,}000$ independently generated samples.

\vspace{-0.3cm}
\subsection{Detection Performance}
For cGN+AWGN, CVAE-MSE attains the earliest rise and saturation, yielding the smallest SNR needed to reach a target $P_d$. Among latent scores, Hermitian Mahalanobis outperforms KLD. ANMF-FP trails the latent Mahalanobis in this Gaussian setting. Doppler-0 behavior is broadly consistent with the mean, with minor shifts.

For cCGN+AWGN, CVAE-MSE remains best overall. However, ANMF-FP becomes the second-strongest detector under compound Gaussian clutter, surpassing latent Mahalanobis and clearly outperforming KLD.
For several detectors (Mahalanobis, KLD), Doppler~0 is slightly \emph{easier} than the mean (left-shifted curves), whereas VAE-MSE exhibits a very slight degradation at Doppler~0.

On experimental CSIR data, ANMF-FP provides the best performance for the Doppler~0, achieving the lowest SNR at moderate-to-high $P_d$.
On the mean over cells, VAE-MSE ranks first, followed by ANMF-FP and Mahalanobis; KLD is the weakest.
Crucially, on the Doppler-0 cell, VAE-MSE degrades markedly, while Mahalanobis remains significantly more robust (improved $P_d$ at the same SNR), and ANMF-FP retains the lead.
This suggests that (i) classical matched-filtering with Tyler robust covariance estimate (ANMF-FP) handles structured, real clutter particularly well, and (ii) incorporating latent covariance geometry (Mahalanobis) alleviates the specific Doppler-0 failure mode observed with pure reconstruction.

All methods saturate near $P_d\!\approx\!1$ at sufficiently high SNR in synthetic settings, with VAE-MSE reaching saturation first. On real data, ANMF-FP saturates fastest, Mahalanobis catches up steadily, and KLD exhibits delayed growth.

\section{Conclusion}
\label{sec:conclu}
We compared complex latent CVAE scores (Mahalanobis, KLD) with VAE-MSE and ANMF-FP across synthetic and real radar data at fixed $P_{fa}$.

On synthetic (cGN/cCGN) data, VAE-MSE is consistently the most SNR-efficient detector; under compound-Gaussian clutter, ANMF-FP is the strongest non-reconstruction alternative, ahead of latent Mahalanobis, while KLD is the weakest. 
On CSIR data, ANMF-FP is best when considering both the mean and the Doppler-0 cell; if one looks only at the mean (excluding Doppler-0), VAE-MSE slightly leads, but it degrades at Doppler-0 where Mahalanobis is markedly more robust. KLD underperforms in all settings.



\bibliographystyle{IEEEbib}
\bibliography{strings,refs}

@article{Robey1992ACA,
  title={A {CFAR} adaptive matched filter detector},
  author={Robey, F. C. and Fuhrmann, D. R.  and  Kelly, E. J. and  Nitzberg, R.},
  journal={IEEE Transactions on Aerospace and Electronic Systems},
  year={1992},
  volume={28},
  pages={208-216},
  url={https://api.semanticscholar.org/CorpusID:17501018}
}

@ARTICLE{4104190,
  author={Kelly, E. J.},
  journal={IEEE Transactions on Aerospace and Electronic Systems}, 
  title={An Adaptive Detection Algorithm}, 
  year={1986},
  volume={AES-22},
  number={2},
  pages={115-127},
  doi={10.1109/TAES.1986.310745}}

@ARTICLE{301849,
  author={Scharf, L. L. and Friedlander, B.},
  journal={IEEE Transactions on Signal Processing}, 
  title={Matched subspace detectors}, 
  year={1994},
  volume={42},
  number={8},
  pages={2146-2157},
  doi={10.1109/78.301849}}

@article{article,
author = {LeCun, Y. and Bengio, Y. and Hinton, G.},
year = {2015},
month = {05},
pages = {436-44},
title = {Deep Learning},
volume = {521},
journal = {Nature},
doi = {10.1038/nature14539}
}

@InProceedings{Adam,
  author    = {Kingma, D. and Ba, J.},
  booktitle = {International Conference on Learning Representations (ICLR)},
  title     = {Adam: A Method for Stochastic Optimization},
  year      = {2015},
  address   = {San Diega, CA, USA},
  optmonth  = {12},
}

@article{Kingma_2019,
   title={An Introduction to Variational Autoencoders},
   volume={12},
   ISSN={1935-8245},
   url={http://dx.doi.org/10.1561/2200000056},
   DOI={10.1561/2200000056},
   number={4},
   journal={Foundations and Trends® in Machine Learning},
   publisher={Now Publishers},
   author={Kingma, D. P. and Welling, M.},
   year={2019},
   pages={307–392} }

@INPROCEEDINGS{marimont2020,
  author={Marimont, S. N. and Tarroni, G.},
  booktitle={2021 IEEE 18th International Symposium on Biomedical Imaging (ISBI)}, 
  title={Anomaly Detection Through Latent Space Restoration Using Vector Quantized Variational Autoencoders}, 
  year={2021},
  volume={},
  number={},
  pages={1764-1767},
  doi={10.1109/ISBI48211.2021.9433778}}

@INPROCEEDINGS{MSWPNM2021,
  author={Mitiche, I. and Salimy, A. and Werner, F. and Boreham, P. and Nesbitt, A. and Morison, G.},
  booktitle={2021 29th European Signal Processing Conference (EUSIPCO)}, 
  title={{OODCN}: Out-Of-Distribution Detection in Capsule Networks for Fault Identification}, 
  year={2021},
  volume={},
  number={},
  pages={1686-1690},
  doi={10.23919/EUSIPCO54536.2021.9615946}}

@INPROCEEDINGS{BS2023,
  author={Bukhsh, Z. and Saeed, A.},
  booktitle={2023 IEEE International Conference on Acoustics, Speech and Signal Processing (ICASSP)}, 
  title={On Out-of-Distribution Detection for Audio with Deep Nearest Neighbors}, 
  year={2023},
  volume={},
  number={},
  pages={1-5},
  doi={10.1109/ICASSP49357.2023.10094846}}

@INPROCEEDINGS{complexreccurrxie,
  author={Xie, Y. and Arildsen, T. and Tan, Z.-H.},
  booktitle={International Joint Conference on Neural Networks (IJCNN)}, 
  title={Complex Recurrent Variational Autoencoder for Speech Resynthesis and Enhancement}, 
  year={2024},
  volume={},
  number={},
  pages={1-7},
  doi={10.1109/IJCNN60899.2024.10650194}}

@article{barrachina:hal-03771786,
  TITLE = {{Comparison Between Equivalent Architectures of Complex-valued and Real-valued Neural Networks - Application on Polarimetric {SAR} Image Segmentation}},
  AUTHOR = {Barrachina, J. A. and Ren, C. and Morisseau, C. and Vieillard, G. and Ovarlez, J.-P.},
  URL = {https://hal.science/hal-03771786},
  JOURNAL = {{Journal of Signal Processing Systems}},
  PUBLISHER = {{Springer}},
  YEAR = {2022},
  MONTH = Jul,
  DOI = {10.1007/s11265-022-01793-0},
  HAL_ID = {hal-03771786},
  HAL_VERSION = {v1},
}

@techreport{an2015variational,
  title        = {Variational Autoencoder based Anomaly Detection using Reconstruction Probability},
  author       = {An, J. and Cho, S.},
  institution  = {SNU Data Mining Center},
  address      = {Seoul National University},
  year         = {2015},
  url          = {http://dm.snu.ac.kr/static/docs/TR/SNUDM-TR-2015-03.pdf}
}

@inproceedings{lee2018simple,
  title     = {A Simple Unified Framework for Detecting Out-of-Distribution Samples and Adversarial Attacks},
  author    = {Lee, K. and Lee, K. and Lee, H. and Shin, J.},
  booktitle = {Advances in Neural Information Processing Systems (NeurIPS)},
  year      = {2018},
  url       = {https://papers.nips.cc/paper_files/paper/2018/hash/abdeb6f575ac5c6676b747bca8d09cc2-Abstract.html}
}

@inproceedings{sohn2015cvae,
  title     = {Learning Structured Output Representation using Deep Conditional Generative Models},
  author    = {Sohn, K. and Lee, H. and Yan, X.},
  booktitle = {Advances in Neural Information Processing Systems (NIPS)},
  year      = {2015}
}

@inproceedings{trabelsi2018deepcomplex,
  title     = {Deep Complex Networks},
  author    = {Trabelsi, C. and Bilaniuk, O. and Zhang, Y. and Serdyuk, D. and Subramanian, S. and Santos, J. F. and Mehri, S. and Rostamzadeh, N. and Bengio, Y. and Pal, C. J.},
  booktitle = {International Conference on Learning Representations (ICLR)},
  year      = {2018},
  url       = {https://openreview.net/forum?id=H1T2hmZAb},
  eprint    = {1705.09792},
  archivePrefix = {arXiv}
}

@article{rouzoumka2025complexvae,
  title={{Complex-Valued Variational Autoencoders for Radar Detection in Joint Compound Gaussian Clutter and Thermal Noise}},
  author={Rouzoumka, Y. A. and Terreaux, E. and Morisseau, C. and Ovarlez, J.-P. and  Ren, C.},
  journal={EUSIPCO},
  year={2025}
}

@article{mahalanobis1936,
  title   = {On the Generalised Distance in Statistics},
  author  = {Mahalanobis, P. C.},
  journal = {Proceedings of the National Institute of Sciences of India},
  volume  = {2},
  pages   = {49--55},
  year    = {1936}
}

@INPROCEEDINGS{9053206,
  author={Hou, Y. and Chen, Z. and Wu, M. and Foo, C.-S. and Li, X. and Shubair, R. M.},
  booktitle={ICASSP 2020 - 2020 IEEE International Conference on Acoustics, Speech and Signal Processing (ICASSP)}, 
  title={Mahalanobis Distance Based Adversarial Network for Anomaly Detection}, 
  year={2020},
  volume={},
  number={},
  pages={3192-3196},
  doi={10.1109/ICASSP40776.2020.9053206}}

@INPROCEEDINGS{nguyen2024vae,
  author={Nguyen, H. H.  and Nguyen, C. N.  and Dao, X. T.  and Duong, Q. T.  and Thi Kim, D. P.  and Pham, M.-T. },
  booktitle={ICCE 2024 - IEEE Tenth International Conference on Communications and Electronics}, 
  title={Variational Autoencoder for Anomaly Detection: A Comparative Study}, 
  year={2024},
  pages={}}

@inproceedings{nakashika20,
  title     = {Complex-Valued Variational Autoencoder: A Novel Deep Generative Model for Direct Representation of Complex Spectra},
  author    = {Nakashika, T.},
  year      = {2020},
  booktitle = {Interspeech 2020},
  pages     = {2002--2006},
  doi       = {10.21437/Interspeech.2020-1964},
  issn      = {2958-1796},
}

@INPROCEEDINGS{RouzoumkaICASSP,
  author={Rouzoumka, Y. A. and Terreaux, E. and Morisseau, C. and Ovarlez, J.-P. and Ren, C.},
  booktitle={IEEE International Conference on Acoustics, Speech and Signal Processing (ICASSP)}, 
  title={Out-of-Distribution Radar Detection in Compound
Clutter and Thermal Noise through Variational
Autoencoders}, 
  year={2025},
  volume={},
  number={},
  pages={},
  doi={}
}

\end{document}